\documentclass[%
reprint,
superscriptaddress,
preprintnumbers,
nofootinbib,
amsmath,amssymb,
 aps,
 prl,
]{revtex4-2}

\usepackage{graphicx}
\usepackage{dcolumn}
\usepackage{bm}
\usepackage{hyperref}
\usepackage[mathlines]{lineno}
\usepackage{float}
\usepackage{siunitx}
\DeclareSIUnit\gauss{G}
\def \bI{{\bf I}}
\def \br{{\bf r}}
\def \bk{{\bf k}}
\newcommand{\bea}{\begin{eqnarray}}
\newcommand{\eea}{\end{eqnarray}}

\begin{document}


\title{An ideal Josephson junction in an ultracold two-dimensional Fermi gas} 
\author{Niclas Luick}
\affiliation{Institut f\"ur Laserphysik, Universit\"at Hamburg}
\affiliation{The Hamburg Centre for Ultrafast Imaging, Universit\"at Hamburg}
\author{Lennart Sobirey}
\affiliation{Institut f\"ur Laserphysik, Universit\"at Hamburg}
\affiliation{The Hamburg Centre for Ultrafast Imaging, Universit\"at Hamburg}
\author{Markus Bohlen}
\affiliation{Institut f\"ur Laserphysik, Universit\"at Hamburg}
\affiliation{The Hamburg Centre for Ultrafast Imaging, Universit\"at Hamburg}
\affiliation{Laboratoire Kastler Brossel, ENS-Universit\'e PSL, CNRS, Sorbonne Universit\'{e}, Coll\`{e}ge de France, 24 rue Lhomond, 75005 Paris, France}
\author{Vijay Pal Singh}
\affiliation{Zentrum f\"ur optische Quantentechnologien, Universit\"at Hamburg, Luruper Chaussee 149, 22761 Hamburg, Germany}
\affiliation{The Hamburg Centre for Ultrafast Imaging, Universit\"at Hamburg}
\author{Ludwig Mathey}
\affiliation{Zentrum f\"ur optische Quantentechnologien, Universit\"at Hamburg, Luruper Chaussee 149, 22761 Hamburg, Germany}
\affiliation{The Hamburg Centre for Ultrafast Imaging, Universit\"at Hamburg}
\author{Thomas Lompe}
\email{tlompe@physik.uni-hamburg.de}
\affiliation{Institut f\"ur Laserphysik, Universit\"at Hamburg}
\affiliation{The Hamburg Centre for Ultrafast Imaging, Universit\"at Hamburg}
\author{Henning Moritz}
\affiliation{Institut f\"ur Laserphysik, Universit\"at Hamburg}
\affiliation{The Hamburg Centre for Ultrafast Imaging, Universit\"at Hamburg}

\date{\today}

\begin{abstract}
\noindent The role of reduced dimensionality in high temperature superconductors is still under debate.
Recently, ultracold atoms have emerged as an ideal model system to study such strongly correlated 2D systems.
Here, we report on the realisation of a Josephson junction in an ultracold 2D Fermi gas.
We measure the frequency of Josephson oscillations as a function of the phase difference across the junction and find excellent agreement with the sinusoidal current phase relation of an ideal Josephson junction.
Furthermore, we determine the critical current of our junction in the crossover from tightly bound molecules to weakly bound Cooper pairs.
Our measurements clearly demonstrate phase coherence and provide strong evidence for superfluidity in a strongly interacting 2D Fermi gas.
\end{abstract}

\maketitle
\noindent One of the most striking macroscopic manifestations of quantum mechanics is the DC Josephson effect \cite{josephson1962possible,Anderson1963}, where a phase difference $\phi$ between two superconductors separated by a weak link drives a current $I(\phi)$ without any applied voltage.
For an ideal Josephson junction, this current phase relation takes a sinusoidal form  $I(\phi) = I_{\rm C} \sin(\phi)$ \cite{golubov2004current}, where $I_{\rm C}$ is the maximum supercurrent that can flow through the junction.
This direct connection between the superfluid current and the phase of the macroscopic wave function makes Josephson junctions a powerful tool for probing properties of superconductors, providing e.g. clear evidence for the \textit{d}-wave symmetry of the order parameter in cuprate superconductors \cite{tsuei2000pairing}.
\begin{figure*} [t]
    \center
    \includegraphics[width = 150mm]{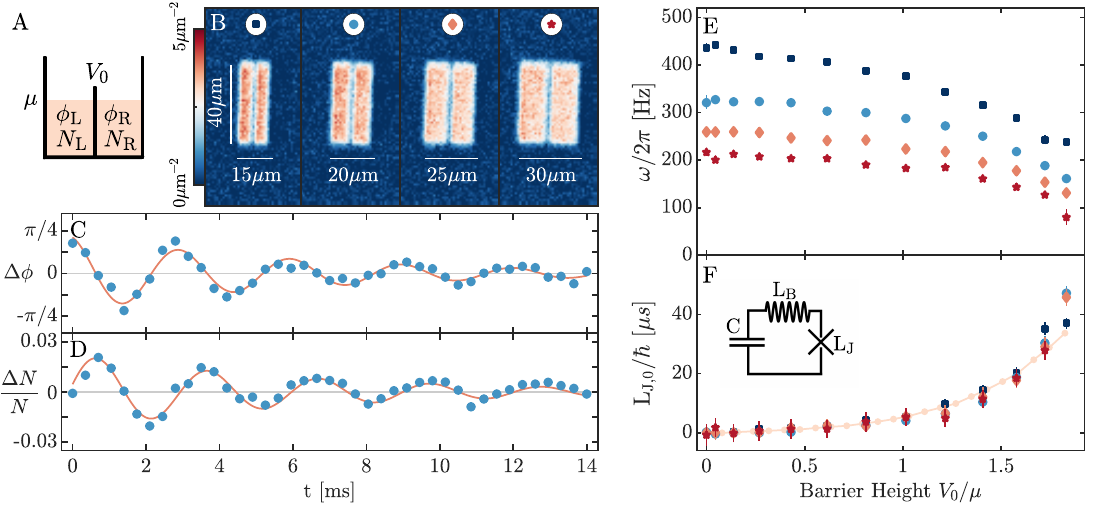}
    \caption{\label{fig:barrier_scan}
    \textbf{Josephson oscillations in a homogeneous 2D Fermi gas.}
    \textbf{(A)} Sketch of a Josephson junction consisting of two Fermi gases with chemical potential $\mu$, particle numbers $N_{\rm L}$, $N_{\rm R}$ and phases $\phi_{\rm L}$, $\phi_{\rm R}$  separated by a tunnelling barrier with height $V_0$. \textbf{(B)} Absorption images of cold atom Josephson junctions. 
The width of the barrier is held fixed at a waist of $w=0.81(6) \mu$m, while the size $l_{\perp}$ of the system is increased. \textbf{(C, D)} Time evolution of the phase difference $\Delta \phi$ (C) and relative particle number difference $\Delta N/N$ (D) between the left and right side of the box after imprinting a relative phase difference of $\phi_0\approx\pi/4$. 
The red lines represent a damped sinusoidal fit.
\textbf{(E)} Oscillation frequency as a function of barrier height $V_0$ for different system sizes (symbols as in (B)), where the error bars denote the 1$\sigma$ fit error. 
The inductance $L_{\rm B}$ and capacitance $C$ of the bulk system are proportional to the length $l_{\perp}$ of the box and therefore the oscillation frequency decreases with increasing system size for $V_0 = 0$.
For nonzero values of $V_0$, the barrier adds a nonlinear Josephson inductance $L_{\rm J}$ to the system and the oscillation frequency decreases as a function of barrier height.   
\textbf{(F)} Josephson inductance $L_{\rm J,0}(V_0)$ extracted from the frequency measurements using an LC circuit model.
The Josephson inductances for all system sizes collapse onto a single curve, which shows that the inductance of the junction depends only on the height of the barrier and validates our LC circuit model. We obtain the calibration of the barrier height $V_0$ by matching the data to a full numerical simulation (dotted line) \cite{SM}. 
The data are obtained by averaging 20 (B), 42 (C), 130 (D) and 7 (E, F) individual measurements.}
\end{figure*}

Recently, ultracold quantum gases have been established as ideal model systems to study such strongly correlated two-dimensional (2D) fermionic systems \cite{Frohlich11,Sommer12,makhalov2014ground,Ong15,fenech2016thermodynamics,Mitra16,mazurenko2017cold,levinsen2015strongly}.
However, although pair condensation of fermions has been reported \cite{Ries15}, fermionic superfluidity in 2D has not been directly observed. 
Here, we use a Josephson junction to unambiguously show phase coherence and provide strong evidence for superfluidity in an ultracold 2D Fermi gas.
Josephson junctions have already been extensively studied in ultracold quantum gases \cite{cataliotti2001josephson,albiez2005direct,levy2007ac,leblanc2011dynamics,betz2011two,spagnolli2017crossing,valtolina2015josephson,burchianti2018connecting,ryu2013experimental}, but the ideal sinusoidal current phase relation that directly links the phase difference to the supercurrent across the junction \cite{watanabe2009critical,spuntarelli2007josephson,ancilotto2009dc} has not been observed \cite{eckel2014interferometric}. 
In this work, we first confirm that our junction follows an ideal current phase relation.
This implies that the current across the junction is a supercurrent that is driven by the phase difference between two superfluids.
We then proceed to measure the evolution of the critical current of the junction as a function of interaction strength and thereby realise a probe for 2D superfluidity in the crossover from tightly bound molecules to weakly bound Cooper pairs.

For our experiments we use a homogeneous Fermi gas of $^6$Li atoms in a spin-balanced mixture of the lowest two hyperfine states, trapped in a box potential \cite{hueck2018two,SM}.
A strong vertical confinement with trap frequency $\omega_{\rm z}/2\pi = 8.8(2)\,{\rm kHz}$ ensures that the gas is kinematically 2D with the chemical potential $\mu$ and temperature $T$ being smaller than the level spacing $\hbar \omega_{\rm z}$, where $\hbar$ is the reduced Planck constant.
We create a Josephson junction by using a narrow repulsive potential barrier with a $1/e^2$ waist of $w=0.81(6)\,\mu{\rm m}$ to split the system into two homogeneous 2D pair condensates connected by a weak link (Figs. \ref{fig:barrier_scan}, A and B).
We imprint a relative phase $\phi_0$ between the two sides of the junction by illuminating one half of the system with a spatially homogeneous optical potential for a variable time between 0 and 20 $\mu$s \cite{SM}.
We then let the system evolve for a time $t$ and extract the population imbalance $\Delta N = (N_{\rm L}- N_{\rm R})$ and the phase difference $\phi$  between the two sides using either in situ or time of flight imaging. 
A typical Josephson oscillation of a molecular condensate at a magnetic field of $B=731\,$G \cite{interaction} and a barrier height of $V_0/\mu = 1.08(5)$ featuring the characteristic $\pi/2$ phase shift between imbalance and phase is shown in Figs. \ref{fig:barrier_scan}, C and D.
The oscillations are weakly damped with a relative damping of $\Gamma/\omega = 0.07$, which according to a full numerical simulation of our system can be explained by phononic excitations in the bulk and the nucleation of vortex-antivortex pairs in the junction (Fig. S\,\ref{fig:temperature}) \cite{lowdamping}.

To understand these Josephson oscillations, we use a simple circuit model commonly used to describe superconducting Josephson junctions \cite{lee2013analogs,eckel2016contact,burchianti2018connecting}.
In this model, we describe our junction as a nonlinear Josephson inductance $L_{\rm J}$ which is connected in series to a linear bulk inductance $L_{\rm B}$ and a capacitance $C$ (Fig. \ref{fig:barrier_scan}F), where the bulk inductance $L_{\rm B}$ characterises the inertia of the gas and the capacitance $C$ its compressibility. 
For vanishing Josephson inductance, the model reduces to a linear resonator with frequency $\omega_s = 1/\sqrt{L_{\rm B} C} = 2\pi v_s/2l_{\perp}$, which corresponds to the frequency of a sound mode propagating with the speed of sound $v_s$ across the length $l_{\perp}$ of the system.
Introducing a barrier with height $V_0$ adds a nonlinear inductance $L_{\rm J}$ to the system and reduces the oscillation frequency $\omega$.
Owing to the nonlinearity of the current phase relation, this $L_{\rm J}$ depends on the phase difference $\phi(t)$ across the junction, but for small phase excitations there is a linear regime where  $L_{\rm J}(\phi(t))$ can be approximated by a time-independent Josephson inductance $L_{\rm J,0}$ and the oscillation frequency is given by $\omega = 1/\sqrt{(L_{\rm B} + L_{\rm J,0}) C}$.

To confirm that our physical system is described by this model, we prepare a gas of deeply bound dimers, perform measurements of the oscillation frequency in the linear regime as a function of the barrier height for different system sizes (Fig. \ref{fig:barrier_scan}E), and extract the Josephson inductance $L_{\rm J,0}$ (Fig. \ref{fig:barrier_scan}F).
Because our system has a uniform density, the bulk inductance is given by the simple expression $L_{\rm B} = 8ml_\perp/\pi^2 n l_{||}$, where $n$ is the density per spin state, $m$ is the mass of a $^6$Li atom, and $l_\perp$ ($l_{||}$) is the diameter of the box perpendicular (parallel) to the barrier \cite{SM}.
Consequently, the Josephson inductance $L_{\rm J,0}(\omega) = L_{\rm B} (\omega_s^2/\omega^2-1)$ can be extracted from the frequency difference between the Josephson oscillations and the sound mode.
Whereas the oscillation frequency is strongly dependent on the size of the box owing to the change in the bulk inductance $L_{\rm B}$ and the capacitance $C$, the measured Josephson inductance $L_{\rm J,0}$ should depend only on the coupling between the two reservoirs. 
As can be seen from Fig.\,\ref{fig:barrier_scan}\,F, all measurements of $L_{\rm J,0}$ versus barrier height collapse onto a single curve regardless of the system size, which confirms that our Josephson junction can be described by an LC circuit model.
For the barrier heights used in our experiments we also find very good agreement with a full numerical simulation of our system  \cite{SM}.

\begin{figure} [t]
    \center
    \includegraphics[width = 75mm]{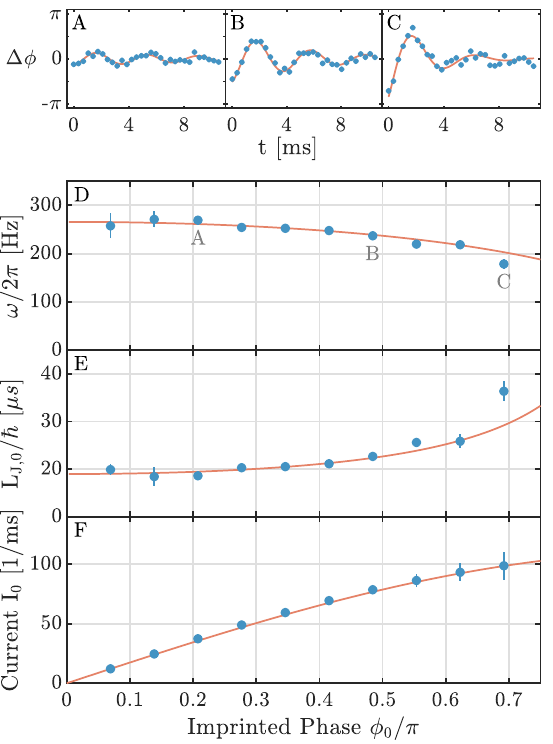}
    \caption{\label{fig:imprinting_scan}
    \textbf{Current phase relation.}
    Josephson oscillations through a tunnelling barrier with height $V_0/\mu=1.51(8)$ at initial phase imprints of \textbf{(A)} $\phi_0 = 0.14 \pi$, \textbf{(B)} $0.42 \pi$ and \textbf{(C)} $0.62 \pi$.
    The amplitude of the oscillations increases for stronger phase imprints, whereas the frequency is reduced.
    \textbf{(D)} Oscillation frequency as a function of imprinted phase, where the error bars denote the 1$\sigma$ fit error.
    \textbf{(E)} Inductance of the junction calculated from the measured oscillation frequencies.
    \textbf{(F)} Effective current $I_0$ through the junction obtained by performing a Riemann sum over the measured values of $L_{\rm J,0}$ shown in (E) according to $\partial I/\partial \phi = \hbar/L_{\rm J}$ \cite{SM}.
    Our data are in excellent agreement with the rescaled current phase relation $I_0 = 2I_{\rm C}\sin(\phi_0/2)$ expected for an ideal Josephson junction (red solid lines), where the initial slope $I_{\rm C}$ is determined from the first three data points. 
    Each data point in (A, B, C) is obtained by averaging 20 individual measurements.
    } 
\end{figure}

Next, we probe the fundamental property of Josephson junctions: the nonlinearity of the current phase relation \cite{eckel2014interferometric,golubov2004current}.
For large phase excitations, the nonlinear current phase relation leads to anharmonic oscillations with an increased oscillation period.
Our ability to imprint arbitrary phase differences $\phi_0$ across the barrier enables us to measure this reduction of the fundamental frequency $\omega(\phi_0)$ as a probe of the nonlinearity (Fig. \ref{fig:imprinting_scan}).
To extract the nonlinear response of the current from our measurements of $\omega(\phi_0)$, we first calculate $L_{\rm J,0}(\omega(\phi_0))$ and then apply the relation $\partial I/\partial \phi = \hbar/L_{\rm J}$ to $L_{\rm J,0}(\phi_0)$ to obtain an effective current $I_0(\phi_0)$.
For an ideal Josephson junction, $I_0$ follows a rescaled current phase relation $I_0(\phi_0) \approx 2I_{\rm C}\sin(\phi_0/2)$  \cite{SM}.
We find that our measurement is in excellent agreement with this current phase relation, indicating that our junction is an ideal Josephson junction \cite{eckel2014interferometric,golubov2004current,idealjunction}.
This implies that the current across the junction is indeed a supercurrent, driven by the phase difference between two superfluids.

\begin{figure} [t]
    \center
    \includegraphics[width = 75mm]{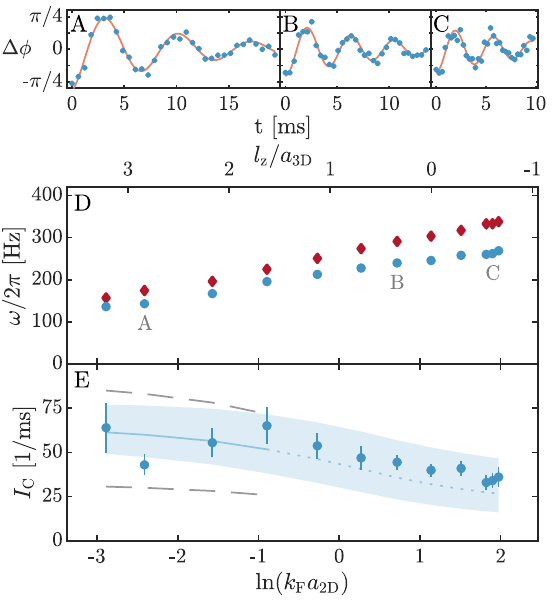}
    \caption{\label{fig:interaction_scan}
    \textbf{Interaction dependence of the critical current.}
    Josephson oscillations for interaction strengths of \textbf{(A)} ${\rm ln}(k_{\rm F} a_{\rm 2D}) = -2.4$, \textbf{(B)} ${\rm ln}(k_{\rm F} a_{\rm 2D}) = 0.7$ and \textbf{(C)} ${\rm ln}(k_{\rm F} a_{\rm 2D}) = 1.9$, where $k_{\rm F}$ is the Fermi wave vector and $a_{\rm 2D}$ is the 2D scattering length as defined in \cite{Ries15,SM}. The measurements are performed in the linear regime with constant density $n=1.21(9) \mu $m$^{-2}$ and relative barrier height $V_0/\mu=1.4(2)$. \textbf{(D)} Oscillation frequency for sound (red diamonds) and Josephson (blue dots) oscillations as a function of the 2D interaction parameter ${\rm ln}(k_{\rm F} a_{\rm 2D})$. The frequency increase of the bare sound mode when going from the molecular to the BCS regime reflects the interaction dependence of the chemical potential. \textbf{(E)} Critical current of the junction extracted from the frequency difference between the sound mode and the Josephson oscillations. The error bars denote the 1$\sigma$ fit error.
    The blue line is the critical current $I_C\propto n_c t_{\bk = 0}$ calculated for a condensate fraction of $n_c/n = 0.72$ and a tunnelling amplitude $t_{\bk = 0}$ obtained from a mean field calculation of the transmission through the barrier \cite{SM}. 
    To calculate the tunnelling amplitude we approximate our junction with a rectangular barrier with a width $b=0.81 \, \mu\rm{m}$, which is a reasonable approximation for the Gaussian barrier used in the experiment.
    The shaded region denotes the systematic uncertainty resulting from the $15 \%$ uncertainty in $V_0/\mu$. 
    The dashed grey lines indicate the upper $(T=0)$ and lower $(T=T_{\rm c})$ bound for the critical current obtained from our theory.
    Although it is unclear how far into the strongly correlated regime our bosonic theory is quantitatively accurate \cite{SM}, it reproduces the qualitative behaviour of our data across the entire BEC-BCS crossover. 
    Each data point in (A, B, C) is obtained by averaging 42 individual measurements.} 
\end{figure}

Following this result, we can now use our Josephson junction as a probe for 2D superfluidity in the strongly correlated regime.
We observe Josephson oscillations over a wide range of interaction strengths, indicating the presence of superfluidity in the entire crossover from tightly bound molecules to weakly bound Cooper pairs (Fig. \ref{fig:interaction_scan}).
To quantify the effect of interactions on our system we extract the critical current $I_{\rm C}$ from the frequency of the Josephson oscillations.
Because for a fixed barrier height $V_0$ the change in the critical current would be dominated by the interaction dependence of the chemical potential, we instead maintain a constant $V_0/\mu = 1.4(2)$ by adjusting the barrier height $V_0$ for each interaction strength according to a reference measurement of the equation of state (Fig. S\ref{FigEOS}).
We observe that, within the uncertainty of our measurement, the critical current stays nearly constant with a tendency towards smaller values of $I_{\rm C}$ when approaching the BCS (Bardeen-Cooper-Schrieffer) side of the resonance.
Although there is currently no theory available that quantitatively describes a 2D Josephson junction in the whole BEC-BCS crossover, in the bosonic limit we can calculate the critical current from the condensate density $n_c$ and the overlap of the condensate wave functions \cite{SM,zaccanti2019critical}.
We use this theory to determine the condensate fraction from the measured critical current for interaction strengths ${\rm ln}(k_{\rm F} a_{\rm 2D}) \leq -0.9$ and obtain $n_c/n = 0.72(8)_{stat.}\left(^{+0.1}_{-0.2}\right)_{sys.}$, where the systematic error arises from the $15 \%$ uncertainty in $V_0/\mu$.
For our homogeneous 2D system, Berezinskii-Kosterlitz-Thouless theory relates the condensate fraction $n_c/n \propto L^{-\eta}$ to the algebraic decay of phase coherence over the finite size $L$ of the box, where $\eta \propto T/n_s$ is the algebraic scaling exponent\cite{hadzibabic2011two,prokof2018algebraic}.
A measurement of the critical current as a function of system size can therefore be used to extract the algebraic scaling exponent and the superfluid density $n_s$ as recently suggested in \cite{singh2020josephson}.

Our homogeneous 2D Fermi gas provides an excellent starting point to study the influence of reduced dimensionality on strongly correlated superfluids in the crossover between two and three dimensions.
The unique combination of reduced dimensionality, uniform density, low entropy and high-resolution imaging makes our system a perfect platform to observe exotic phases such as the elusive Fulde-Ferrell-Larkin-Ovchinnikov state \cite{kinnunen2018fulde}.
Finally, our system is ideally suited to investigate whether periodic driving of Josephson junctions can strongly enhance coherent transport, as suggested by experiments with THz-driven cuprate superconductors \cite{hu2014optically,okamoto2017transiently}.

\section*{Acknowledgments}
\noindent We thank K. Hueck and B. Lienau for their contributions during earlier stages of the experiment, T. Enss, A. Recati and M. Zaccanti for stimulating discussions and G. Roati and F. Scazza for careful reading of the manuscript and valuable suggestions regarding the interpretation of Fig. 3.
\textbf{Funding:} This work was supported by the European Union's Seventh Framework Programme (FP7/2007-2013) under grant agreement No. 335431 and by the DFG in the framework of SFB 925 and the excellence clusters 'The Hamburg Centre for Ultrafast Imaging'- EXC 1074 - project ID 194651731 and 'Advanced Imaging of Matter' - EXC 2056 - project ID 390715994.
M. Bohlen acknowledges support by Labex ICFP of \'{E}cole Normale Sup\'{e}rieure Paris.
\textbf{Author contributions:} N.L. and L.S. performed the experiments and data analysis with support from M.B. and T.L..
V.P.S. and L.M. developed numerical and analytical models and contributed to the interpretation of our experimental data. 
N.L. and T.L. wrote the manuscript and L.S. created the figures with input from all authors. 
T.L. and H.M. supervised the project. 
All authors contributed to the discussion and interpretation of our results.
\textbf{Competing interests:} The authors declare no competing interests.
\textbf{Data and materials availability:} All data presented in this paper and simulation scripts are deposited in \cite{luick_niclas_2020_3744797,luick_niclas_2020_3786298}.



\clearpage
\pagebreak
\begin{center}
\textbf{\large Supplementary materials}
\end{center}
\setcounter{figure}{0}
\renewcommand{\figurename}{Fig.\,S}


\subsection*{Preparation of homogeneous 2D Fermi gases}

We prepare our homogeneous 2D Fermi gas following the scheme described in \cite{hueck2018two}. 
We start by evaporatively cooling a spin mixture of $^6{\rm Li}$ atoms in the $\left|F=1/2, m_F =1/2\right>$ and $\left|F=1/2, m_F =-1/2\right>$ hyperfine states in a highly elliptical optical dipole trap at a magnetic field close to the 832\,G Feshbach resonance of $^6{\rm Li}$.
We then ramp to a magnetic field of 730\,G and project our box potential onto the atoms using a digital micromirror device (DMD\footnote{Texas Instruments DLP6500FYE}) illuminated with blue-detuned ($\lambda=532 \, {\rm nm}$) light, which we refer to as DMD 1.
Additionally, we briefly ramp up a second DMD (DMD 2), also illuminated with $532 \, {\rm nm}$ light, that covers a larger area to push away residual atoms still trapped outside of the box.
Finally, we load the atoms into a single node of an optical standing wave potential with a lattice spacing of approximately $3\,\mu{\rm m}$ and a trap frequency of $\omega_z = 2 \pi \cdot 8.8(2)\,{\rm kHz}$ and thereby bring the atoms into the 2D-regime.
For all measurements, the chemical potential is well below the trap frequency ($\mu < 0.7 \, \hbar \omega_z$) and we can therefore parametrise the interaction strength by an effective 2D scattering length $a_{\rm 2D}=l_{\rm z}\sqrt{\pi/0.905} \exp(-\sqrt{\pi/2} \cdot l_{\rm z}/a_{\rm 3D})$ \cite{petrov2001interatomic}, where $l_{\rm z} = \sqrt{\hbar/m \omega_{\rm z}}$ is the harmonic oscillator length and $a_{\rm 3D}$ is the 3D scattering length.

We note that performing thermometry of our homogeneous Fermi gas is challenging, since, in contrast to harmonic traps, there is no low density region where the gas is thermal. 
This makes it very difficult to observe and fit the thermal fraction of the cloud. 
We obtain an estimate of the temperature of the system by performing a time of flight measurement after DMD 1 has been ramped on and the atoms have been loaded into the lattice, but without pushing away the atoms outside the box with DMD 2. 
This measurement yields a temperature of $T/T_{\rm F} \approx 0.03$, where $T_{\rm F}= E_{\rm F}/k_{\rm B}$ is the Fermi temperature of a system with Fermi energy $E_{\rm F}=\hbar^2 k_{\rm F}^2/2m = \hbar^2 4\pi n/2m$ using the density $n$ inside the box potential. 
However, it is \textit{a priori} unclear whether the atoms inside and outside the box potential are fully thermalised, so a better method to perform thermometry is to measure the density equation of state $\mu(n,T)$ of a molecular condensate and compare it to a full numerical simulation (Fig. S4).
This method yields the same temperature of $T/T_{\rm F} \approx 0.03$ as the time of flight measurement.

\subsection*{Generation of arbitrary potentials}

\begin{figure} [t]
    \center
    \includegraphics[width = 75mm]{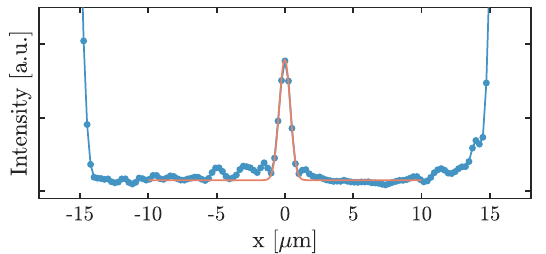}
    \caption{\label{fig:barrier} \textbf{Calibration of the barrier.}
    Line sum through an image of a box potential with a barrier in the center. The box has a width of 300 pixels on the DMD, which corresponds to a box size of $30\,\mu{\rm m}$ in the plane of the atoms. The barrier has a width of 4 pixels, which is broadened by the finite resolution of the imaging system used to project the image onto the atoms. From a Gaussian fit to the barrier (red line), we determine a $1/e^2$ waist of $w = 0.81(6) \mu {\rm m}$.}
\end{figure}

To create the box potential and the tunnelling barrier, we image DMD 1 directly onto the atoms using a high resolution microscope.
The DMD has a pixel size of $7.56 \, \mu {\rm m}$ and is demagnified by a factor of 75 by the imaging system, so that each DMD pixel has a width of $0.1\,\mu{\rm m}$ in the image plane, which is much smaller than the resolution of the imaging system. 
For narrow barriers with a width $W\lesssim 10$\, pixel, we can therefore adjust the height of the barrier by increasing the width of the barrier on the DMD image.
We characterise the tunnelling barrier by using a second high resolution microscope to image the intensity distribution in the plane of the atoms (Fig. S \ref{fig:barrier}) for all barrier widths $W$ used in our experiments.
From these images, we obtain a calibration of the relative change of the barrier height as a function of $W$ as well as a determination of the barrier width $w = 0.81(6) \, \mu {\rm m}$, which is independent of $W$ for $W \leq 11$ pixel (deviation $< 10 \%$).

\subsection*{Phase control}
\begin{figure} [t]
    \center
    \includegraphics[width = 75mm]{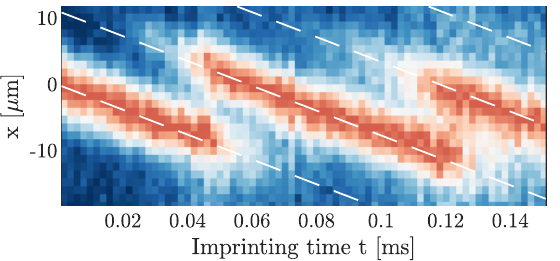}
    \caption{\label{fig:phase}
    \textbf{Calibration of phase imprinting}
    We vary the time for which the imprinting potential is switched on and measure the shift in the position of the central momentum peak. From its periodic displacement, we obtain the height of the optical potential $\Delta V_0=h \cdot 16.0(5)$ kHz. Each column represents a slice through the momentum distribution.
    The data shown is obtained by averaging over 38 realisations.}
\end{figure}

To imprint a relative phase between the two sides of our Josephson junction, we briefly apply an optical potential $\Delta V_{0}$ to the condensate in one of the reservoirs, which advances its phase by $\Delta V_{0} t/ \hbar $.
The time $t$ is much shorter than the Fermi time $h/E_{\rm F}$, which ensures an almost pure phase excitation.
The spatially homogeneous optical potential is created by DMD 2 and imaged onto the atom plane. We perform matter wave imaging \cite{hueck2018two,murthy2014matter} and observe the relative phase difference between the reservoirs in the displacement of the pair condensation peak.
We calibrate this procedure by measuring the periodic displacement of the momentum peak as a function of the time for which the imprinting potential is switched on. 
From the measurement shown in Fig. S \ref{fig:phase}, we obtain $\Delta V_{0}=h \cdot 16.0(5)$ kHz for the potential height\footnote{We define $\Delta V_{0}$ as the potential experienced by a pair of atoms, which has two times the polarisability and therefore experiences twice the optical dipole potential as a single atom.} and a displacement $\Delta x = 5.5(3) \mu {\rm m}/\pi$.

\subsection*{LC Circuit Model}
We use an electrical circuit model similar to the one used in \cite{lee2013analogs,eckel2016contact,burchianti2018connecting} to model the dynamics of our Josephson junction.
We describe our system as a capacitance $C$ and a bulk inductance $L_{\rm B}$ connected in series to a Josephson junction, which is modelled as a nonlinear inductance ${L}_{\rm J}(\phi)$, see Fig. 1F.
In this circuit, the current $I=\frac{1}{2}\frac{d(\Delta N)}{d t}$ is the instantaneous particle current across the junction determined by the change in the particle number imbalance $\Delta N = N_{\rm L} - N_{\rm R}$.
The voltage over the capacitor is given by $U_{\rm C}= \Delta N/2 C$, where $\Delta N/2$ corresponds to the charge of the capacitor.
The voltage across the junction $U_{\rm J} = L_{\rm J} \frac{dI}{dt}$ is related to the phase difference $\phi$ via the Josephson-Anderson relation $U_{\rm J}=\hbar \frac{d \phi}{dt}$ and hence the junction has an inductance of $L_{\rm J}(\phi)=\hbar/\frac{dI(\phi)}{d\phi}$.
According to Kirchhoff's law, the voltages across the capacitor, the bulk inductance and the junction add to zero and therefore the LC circuit is described by the differential equation
\begin{equation}\label{eq:eom}
\frac{\Delta N}{C} + \left( L_{\rm B} + L_{\rm J}(\phi) \right) \frac{\partial^2(\Delta N )}{\partial t^2}= 0.
\end{equation} 

\subsubsection*{Linear Regime}
For small phase excitations, $L_{\rm J}(\phi)$ can be approximated by a constant, phase independent inductance $L_{\rm J,0}$ and Eq. \ref{eq:eom} yields harmonic oscillations with frequency $\omega = \frac{1}{\sqrt{(L_{\rm B}+L_{\rm J,0}) C}}$.
For a vanishing barrier, the oscillations correspond to a phononic excitation propagating between the boundaries of the box at the speed of sound $v_{\rm s}$.
In the circuit model, this corresponds to $L_{\rm J}=0$ and the frequency is given by $\omega_s = \frac{1}{\sqrt{L_{\rm B} C}} = 2\pi\frac{v_{\rm s}}{2 l_\perp}$.
Hence, we can calculate $L_{\rm J,0}$ from the ratio of the oscillation frequencies
\begin{equation}\label{eq:lj}
L_{\rm J,0} = L_{\rm B} \left(\frac{\omega_s^2}{\omega^2} - 1\right).
\end{equation}
For our homogeneous box system, the speed of sound $v_{\rm s} = \sqrt{\frac{n}{m} \frac{\partial \mu}{\partial n}}$ and the capacitance $C=\frac{1}{2}\frac{\partial N}{\partial \mu_B}=\frac{1}{4}\frac{\partial N}{\partial \mu}=\frac{1}{8}l_\perp l_{||} \frac{\partial n}{\partial \mu}$ are related to each other by the compressibility $\kappa = \frac{\partial \mu}{\partial n}$, where $\mu_{\rm B} = 2 \mu$ is the chemical potential of a gas of bosonic dimers.
Therefore, we can simply calculate the bulk inductance $L_{\rm B} = 1/\omega_s^2 C= 8m\,l_\perp/\pi^2 n \,l_{||}$ and thereby obtain the Josephson inductance $L_{\rm J,0}$ without using the equation of state $\mu(n)$.

\subsubsection*{Current Phase Relation}
For a phase excitation that is not small, the nonlinearity of the Josephson inductance leads to anharmonic oscillations.
This nonlinear oscillation consists of a down-shifted fundamental frequency $\omega(\phi_0)$ and higher harmonics of this frequency.
One possibility to extract the current phase relation from this anharmonic oscillation would be to obtain the instantaneous current $I(\phi(t))$ and relating it to the corresponding $\phi(t)$. 
However, this approach has the significant drawback that obtaining $I(\phi(t))$ requires numerical differentiation of $\Delta N(t)$, which is extremely sensitive to noise.
Hence, we use the information contained in the shift of the fundamental frequency $\omega(\phi_0)$ to probe the current phase relation.
We do this by extracting $L_{\rm J,0}(\phi_0)$ from $\omega(\phi_0)$ according to Eq. \ref{eq:lj} and then calculating an effective current $I_0(\phi_0) = \int_0^{\phi_0} \frac{\hbar}{L_{\rm J,0}(\phi_0')}\, d\phi_0'$ by performing a Riemann sum over all experimentally determined values of $\frac{\hbar}{L_{\rm J,0}(\phi_0')}$ for which $\phi_0'\le\phi_0$.
While this effective current is different from the instantaneous current, we can still relate the effective current phase relation $I_0(\phi_0)$ and the instantaneous current phase relation $I(\phi)$ by inserting the ideal current phase relation $I(\phi)=I_{\rm C} \sin(\phi)$ into Eq. \ref{eq:eom}.
In principle, $I_0(\phi_0)$ can be found by solving Eq. \ref{eq:eom} numerically, but it is instructive to consider a simplified case which can be solved analytically. 
If we assume that the dynamics of the system is dominated by the barrier $(L_{\rm J} \gg L_{\rm B})$, we can neglect the bulk inductance $L_{\rm B}$ in the LC circuit and write $\hbar\dot \phi + \frac{\Delta N}{2C}= 0$.
Differentiating this equation, we get
\begin{equation}
\ddot{\phi}+\frac{I_{\rm C}}{\hbar C}\sin(\phi) = 0 \, ,
\end{equation}
which is equivalent to the equation of motion of a mathematical pendulum. 
To first order, the oscillation frequency is given by $\omega(\phi_0)^2 \approx \frac{I_{\rm C}}{\hbar C} (1-\phi_0^2/8) \approx \frac{I_{\rm C}}{\hbar C} \cos \frac{\phi_0}{2}$
and we can extract the corresponding inductance
\begin{equation}
L_{\rm J,0}(\phi_0) \approx L_{\rm J,0,\phi_0 \rightarrow 0}\left(\frac{\omega_{\phi_0 \rightarrow 0}}{\omega(\phi_0)}\right)^2  \approx \frac{\hbar}{I_{\rm C} \cos{(\phi_0 / 2)}}\,.
\end{equation}
After integration, we get a simple rescaled expression for the effective current
\begin{equation}\label{eq:IphiApprox}
I_0(\phi_0) \approx 2 I_{\rm C} \sin{\frac{\phi_0}{2}} \,.
\end{equation}
We compare this result with the current $I_0(\phi_0)$ extracted from the numerical solution of Eq. \ref{eq:eom} for a system with $L_{\rm J,0}/L_{\rm B} = 1.3$, which is the value of $L_{\rm J,0}/L_{\rm B}$ for the system that was used for the measurements in Fig. \ref{fig:imprinting_scan}.
We find that for initial phase excitations $\phi_0\lesssim 0.7\pi$, Eq. \ref{eq:IphiApprox} and the numerical solution agree within $2\%$.
Hence, we compare our data to Eq.\,\ref{eq:IphiApprox}.

\subsection*{Numerical simulations}
\begin{figure}[t]
    \begin{center}
        \includegraphics[width=0.75\linewidth]{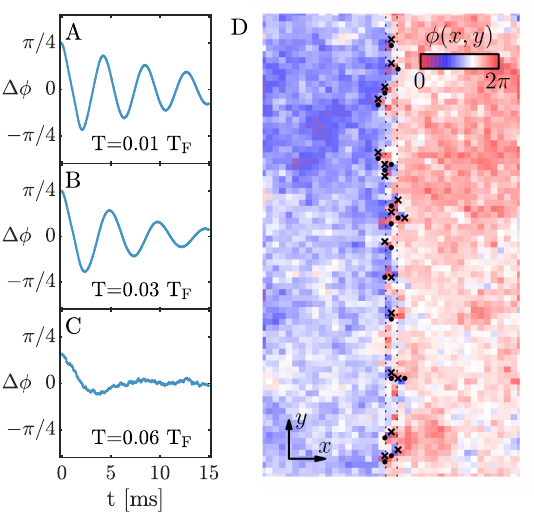}
        \caption{\label{fig:temperature} \textbf{Temperature dependence of Josephson oscillations.}  
        \textbf{(A-C)} Time evolution of the simulated $\Delta \phi$ for $V_0/\mu \approx 1.0$ and a phase imprint of $\pi/4$ at three different temperatures.
        \textbf{(D)} Simulated phase evolution of one sample of the ensemble, $3.9\,$ms after a phase imprint of  $\pi/4$, for $n=2.25\, \mu$m$^{-2}$ and $T/T_{\rm F} \approx 0.03$.
        The barrier height is $V_0/\mu \approx 2$ and its width of $0.85\, \mu$m is denoted by the two vertical dotted lines. 
        The dots and the crosses represent vortices and antivortices, respectively. 
        The box dimensions are $20 \times 40\,\mu$m$^2$.}
    \end{center}
\end{figure}

We simulate the dynamics of a two-dimensional (2D) bosonic Josephson junction with the c-field simulation method that was used in Ref. \cite{singh2017superfluidity}. 
Our homogeneous 2D system is described by the Hamiltonian
\begin{equation}
\label{eq_hamil}
\hat{H}_{0} = \int \mathrm{d}{\bf r} \left[ \frac{\hbar^2}{2M}  \nabla \hat{\psi}^\dagger({\bf r}) \cdot \nabla \hat{\psi}({\bf r})  + \frac{g}{2} \hat{\psi}^\dagger({\bf r})\hat{\psi}^\dagger({\bf r})\hat{\psi}({\bf r})\hat{\psi}({\bf r})\right] ,
\end{equation}
where $\hat{\psi}$ and $\hat{\psi}^\dagger$ are the bosonic annihilation and creation operators, respectively.
The interaction $g$ is given by $g=\tilde{g} \hbar^2/M$, where  $\tilde{g}$ is the dimensionless interaction and $M$ the mass of a $^{6}$Li$_2$ molecule. 
Here, $\tilde{g}$ is determined by $\tilde{g}=\tilde{g}_0/\bigl(1- \frac{\tilde{g}_0}{2\pi} \ln(2.09 k_{\rm F} \ell_d) \bigr)$, with  $\tilde{g}_0= \sqrt{8 \pi} a_s/\ell_d$ \cite{turlapov2017fermi}, where $a_s$ is the molecular s-wave scattering length, $\ell_d= \sqrt{\hbar/(M \omega_z)}$ is the harmonic oscillator length in the transverse direction, and $k_{\rm F}$ is the Fermi wave vector.
Analogous to the experiments, we consider 2D clouds of $^{6}$Li$_2$ molecules confined in a box of dimensions $L_x \times L_y$. 
We discretise the space with a lattice of size $N_x \times N_y$ and a discretisation length $l=0.5\, \mu \mathrm{m}$. 
Within our c-field representation, we describe the operators $\hat{\psi}$ in Eq. \ref{eq_hamil} and the equations of motion by complex numbers $\psi$.
We sample the initial states in a grand canonical ensemble with chemical potential $\mu$ and temperature $T$ via a classical Metropolis algorithm. 
The system parameters, such as the density $n$, $\tilde{g}$, and $T$ are chosen in accordance with the experiments.  
To simulate the Josephson junction we add the term $\mathcal{H}_{ex} = \int \mathrm{d}{\bf r}\, V({\bf r}) n({\bf r})$, where $n({\bf r})$ is the density at the location  ${\bf r}=(x,y)$. 
The barrier potential $V({\bf r})$ is given by 
\begin{equation}\label{eq_pot} 
V({\bf r})  = V_0\exp \bigl(- 2 (x-x_0)^2/w^2 \bigr),
\end{equation}
where $V_0$ is the barrier height and $w$ the width. 
The potential is centered at the location $x_0= L_x/2$. 
We choose $w=0.85\, \mu \mathrm{m}$ and $V_0/\mu$ in the range $0-2$, where we use $\mu = gn$. 
This splits the system in $x$-direction into two equal 2D clouds, which we refer to as the left and right reservoir. 
We then imprint a fixed value of the phase on one of the reservoirs, which creates a phase difference $\Delta \phi= \phi_{\rm L} - \phi_{\rm R}$, where $\phi_{\rm L}$ ($\phi_{\rm R}$) is the mean value of the phase of the left (right) reservoir.
The sudden imprint of phase results in oscillations of $\Delta \phi$ and the density imbalance $\Delta N = N_{\rm L}-N_{\rm R}$, where $N_{\rm L}$ ($N_{\rm R}$) is the number of molecules in the left (right) reservoir.
We analyse the time evolution of $\Delta N$ and $\Delta \phi$ for system parameters close to the ones used in the experiments. 
Fig. S \ref{fig:temperature}\,A-C shows simulations of $\Delta \phi(t)$ at three different temperatures of $T/T_{\rm F} \approx 0.01$, $0.03$, and $0.06$ for $n=2.25\, \mu \mathrm{m}^{-2}$, $\tilde{g}=1.8$ and a system size of $L_x \times L_y = 20 \times 40\, \mu \mathrm{m}^2$.  
The damping of the oscillations increases with temperature.
To quantify this observation, we fit $\Delta \phi (t)$ with a damped sine function $f(t)= A_0 e^{-\Gamma t} \sin(\omega t + \theta)$, where $A_0$ is the amplitude, $\omega$ is the oscillation frequency, $\Gamma$ is the damping, and $\theta$ is the phase shift. 
The determined ratio of  $\Gamma / \omega$ is $0.05$, $0.09$, and $0.45$ for $T/T_{\rm F} \approx 0.01$, $0.03$, and $0.06$, respectively.
As the experimentally observed damping is on the order of $\Gamma / \omega \approx 0.07$, this suggests an experimental temperature on the order of $T/T_{\rm F} \lesssim 0.03$, which is consistent with the results from measurements of the momentum distribution and the equation of state shown in Fig. S \ref{FigEOS}.
To obtain a calibration of the experimental barrier height, we simulate the system for a wide range of barrier heights $V_0$ and match the simulated and extracted Josephson inductances by fitting the calibration factor between the width $W$ of the barrier on the DMD and the simulated barrier height $V_0$.

To understand the mechanism for the thermal damping of the oscillations, we examine the phase evolution of a single sample of our ensemble. 
Figure S\,\ref{fig:temperature}D shows the phase $\phi(x,y)$ at a point in time which is $3.9\,$ms after a phase imprint of $\pi/4$ for the same $n$, $\tilde{g}$ and box size as above, and $T/T_{\rm F} \approx 0.03$. 
At this time the system exhibits distinct values of the mean phase for the left and right reservoir and a strong phase gradient across the barrier.
As expected for a 2D system, the phase is weakly fluctuating within the reservoirs due to thermal phonons.
In addition to the phonons, we identify the nucleation of vortex-antivortex pairs as an additional mechanism of dissipation. 
We calculate the phase winding around a lattice plaquette of size $l\times l$ using $\sum_{\Box} \delta \phi(x,y) = \delta_x\phi(x,y) + \delta_y\phi(x+l,y)+\delta_x\phi(x+l,y+l)+\delta_y\phi(x,y+l)$, where the phase differences between sites is taken to be $\delta_{x/y} \phi(x,y)  \in (-\pi, \pi]$. 
We show the calculated phase winding in Fig. S \ref{fig:temperature}D. 
A vortex and an antivortex are identified by a phase winding of $2\pi$ and $-2\pi$, respectively. The vortex pairs are nucleated mainly inside the barrier in the regions of low densities. 
Both the phonons and vortices lead to the damping of oscillations shown in Fig. S \ref{fig:temperature} \,A-C. 

\subsection*{Equation of state}

\begin{figure}[t]
    \begin{center}
        \includegraphics[width=0.75\linewidth]{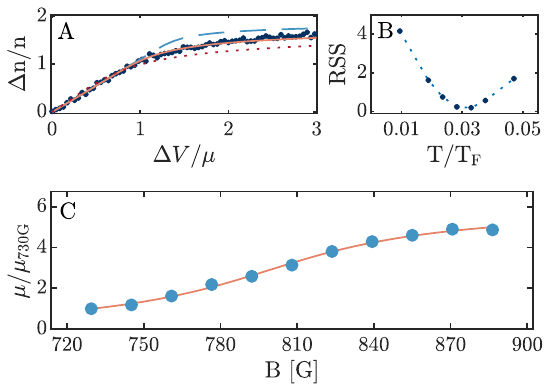}
        \caption{\textbf{Equation of state.}
        \textbf{(A)} Density difference $\Delta n$ (dark blue dots) created by a potential step $\Delta V$, compared to numerical simulations performed at $T/T_{\rm F}=0.019$ (dashed blue line), $T/T_{\rm F}=0.033$ (solid red line) and $T/T_{\rm F}=0.047$ (dotted dark red line). 
        \textbf{(B)} Residual sum of squares between the numerical simulations performed at different temperatures and the measured EOS.
        The dashed blue line is a guide to the eye.
        The best agreement between our measured equation of state and the simulation is achieved at a temperature of $T/T_{\rm F} = 0.03$.
        \textbf{(C)} 
        Relative change of the chemical potential of our system for different magnetic fields, normalized to the chemical potential at a field of $730 \,{\rm G}$. 
        The chemical potential is extracted from the initial slope of the EOS measurements shown in \textbf{(A)}.
        The red line is a heuristic fit we use to keep $V_0/\mu$ constant during our measurement across the crossover (Fig. \ref{fig:interaction_scan}).
        The data shown is obtained by averaging over 7 \textbf{(A)} and 3 \textbf{(C)} realisations.}
        \label{FigEOS}
    \end{center}
\end{figure}

To keep our relative barrier height $V_0/\mu$ constant during measurements over the crossover, we need to measure the chemical potential $\mu$ as a function of interaction strength.
We do this by following the approach established by Ref. \cite{boettcher2016equation}.
We therefore define our chemical potential as $\mu=\mu_0+\epsilon_{\rm B}/2$, where $\mu$ is the chemical potential per atom and the contribution of the two-body binding energy $\epsilon_{\rm B}$ is subtracted from the bare chemical potential $\mu_0$.
We use DMD 2 to introduce a potential offset $\Delta V$ between the two sides of the box and measure the resulting density difference $\Delta n$. 
For sufficiently small temperatures, the Thomas-Fermi approximation predicts $E_{\rm F}=c\cdot\mu$, and we can obtain $\mu/E_{\rm F}=1/c$ from the linear slope of $\Delta n (\Delta V)$.
The relative change in the measured chemical potential for different magnetic fields is shown in Fig. S \ref{FigEOS}\,C. 

To determine the temperature of our system we compare the measured $\Delta n (\Delta V)$ of the bosonic system with interaction strength ${\rm ln}(k_{\rm F} a_{\rm 2D}) = -2.9 $ with simulated curves for different temperatures obtained using the c-field method outlined above. 
The discrepancy between the measurement and the numerical simulations is minimised for a temperature of $T\approx0.03 \, T_{\rm F}$ (Fig. S \ref{FigEOS}A,B).

\subsection*{Calculation of the critical current}

In the following we derive an analytic expression for the critical current $I_{\rm C}$ of our 2D Josephson junction in the bosonic limit including phase fluctuations, motivated by recent work for 3D systems \cite{zaccanti2019critical}.
Generally, we can express the current between the left and right reservoir

\bea
\label{eq:ic1}
\bI &=& - \frac{i}{\hbar} \Big( \sum_{\bk} t_{\bk} (a^{\dagger}_{l}(\bk) a_{r}(\bk) -  a^{\dagger}_{r}(\bk) a_{l}(\bk)    ) \Big)
\eea
via the tunnelling amplitudes $t_{\bk}$ and the bosonic creation and annihilation operators acting on the left and right reservoir. 
In the phase-density representation, neglecting density fluctuations, the annihilation operators are given by
\bea
\label{eq:ic2}
a_{l/r}(\bk) &=& \int \frac{d^{2} r}{\sqrt{A}} \exp(- i \bk \br) \sqrt{n_{l/r}} \exp(i \phi_{l/r} + i \delta \phi_{l/r}(\br)) \, ,
\eea
where $A = L^2$ is the area of a box of size $L$, $n_{l}  \left(n_{r} \right)$ is the density, $\phi_{l} \left(\phi_{r}\right)$ is the phase, and $\delta\phi_{l} \left(\delta \phi_{r}\right)$ is the fluctuation of the phase in the left (right) reservoir.
To calculate the expectation value $\langle \bI \rangle$, we insert Eq. \ref{eq:ic2} in Eq. \ref{eq:ic1} and assume independent Gaussian fluctuations of the phase in both reservoirs $\langle e^{i\delta \phi_{l/r}(\br)} \rangle = e^{-\frac{1}{2}\langle \delta \phi_{l/r}^2 (\br) \rangle}$.
For a 2D system, we can further approximate the phase fluctuations to lowest order as $\langle \delta\phi^{2}_{l/r}(\br) \rangle = \eta \log(L/r_{0})$, where $\eta =  \frac{M k_{B} T}{2 \pi \hbar^{2} n_s} = 2 \frac{n}{n_s} \frac{T}{T_{\rm F}}$ is the algebraic scaling exponent and $r_0\approx \xi$ is a short range cutoff on the order of the system's healing length $\xi$.
To lowest order in $k$, we obtain
\bea
\langle \bI \rangle &=&  \frac{2 n A t_{\bk=0} }{\hbar} \left(\frac{L}{r_0}\right)^{-\eta} \sin\phi \, ,
\eea
where $\phi = \phi_{r} - \phi_{l}$ is the phase difference across the barrier.
This result reproduces the ideal current phase relation $I(\phi) = I_{\rm C} \sin(\phi)$, where the critical current $I_{\rm C}$ is reduced by a factor of $\left(\frac{L}{r_0}\right)^{-\eta}$. 
Therefore, the critical current $I_{\rm C}$ is directly related to the algebraic decay of phase coherence in a 2D superfluid.
Using the the condensate density
\bea
n_c \approx n (L/r_{0})^{-\eta}
\eea
of a finite size 2D gas as defined in ref. \cite{hadzibabic2011two} we finally get the critical current
\bea
\label{eq:theorycurrent}
I_{\rm C} \approx \frac{2 n_c A  t_{\bk=0} }{\hbar} \, .
\eea
We calculate the tunnelling amplitude $t_{\bk=0}$ for a rectangular potential barrier of width $d$ and height $V_B$, centered around $x=0$, with the following mean field ansatz
\bea
\psi(x) &=&  \begin{cases}
    -\frac{1}{\sqrt{L}} \tanh((x+\delta)/(\sqrt{2}\xi))              & x<-d/2\\
    B \exp(-\kappa(x+d/2))               & x>-d/2 \, ,
\end{cases}
\eea
with 
\bea
\delta &=& \frac{d}{2} - \frac{\xi}{\sqrt{2}} \textrm{arcsinh}\Big( \frac{\sqrt{2}}{\kappa \xi}\Big) \, \\
B &=& \frac{1}{\sqrt{L}} \tanh\Big(\frac{1}{2} \textrm{arcsinh}\Big( \frac{\sqrt{2}}{\kappa \xi}\Big)\Big) \,
\eea
where $\xi = \hbar/\sqrt{2 M \mu_B}$ is the healing length for a gas of bosons with mass $M$ and chemical potential $\mu_B$.
Outside the barrier $(x<-d/2)$, $\psi(x)$ is the exact solution to the Gross-Pitaevskii equation.
Inside the barrier $(x>-d/2)$, we obtain the approximative solution by minimising the energy
\bea
E &=& \frac{ B^{2}}{2\kappa} \Big( \frac{\hbar^{2} \kappa^{2}}{2 M} + V_B - \mu_B \Big) + \frac{g}{2} \frac{B^{4}}{4 \kappa} \,
\eea
which yields the characteristic decay exponent $\kappa = \sqrt{k_{0}^{2} + k_{B}^{2}}$ with $k_{0}^{2} = 2 M (V_B-\mu_B)/\hbar^{2}$ and $k_{B}^{2} = M g B^{2}/2\hbar^{2}$.
Using the continuity of the wave function and its derivative at $z = -d/2$ we further get
\bea
\kappa^{2} &=& k_{0}^{2} + \frac{ M g B^{2}}{2\hbar^{2}} = \frac{n }{2 \xi^{2}} \Big(1 - \frac{B^{2}}{n}\Big)^{2} \frac{1}{B^{2}}
\eea
and obtain 
\bea
B^{2} &=& \frac{n}{1 + k_{0}^{2} \xi^{2}  + \sqrt{ 1/2 + 2 k_{0}^{2} \xi^{2} + (k_{0}^{2} \xi^{2})^{2}  } }\\
&=& n \frac{\mu_B}{V + \sqrt{V_B^{2} - \mu_B^{2}/2}} \, .
\eea
Therefore, we finally obtain the tunnelling energy
\bea
t_{\bk=0} &=& \frac{1}{L} \frac{\hbar^{2} \kappa}{m} \frac{\mu_B}{V_B + \sqrt{V_B^{2} - \mu_B^{2}/2}}  \exp(- \kappa d) \, ,
\eea
which we insert into Eq. \ref{eq:theorycurrent} to obtain the critical current $I_{\rm C}$.

We note that Eq. \ref{eq:theorycurrent} which relates the critical current to the condensate density has previously been derived for a 3D system \cite{meier2001josephson} and was 
found to quantitatively describe the behaviour across the 3D BEC-BCS crossover for high barriers $(V\gg \mu)$ \cite{zaccanti2019critical}. 
Our derivation of the critical current extends the validity of this relation in the bosonic limit to $V\gtrsim \mu$ by including the mean field contribution to the tunnelling amplitude inside the barrier.
Following the same reasoning and assumptions given in \cite{zaccanti2019critical}, it seems plausible that Eq. \ref{eq:theorycurrent} is quantitatively accurate beyond the bosonic case discussed above, but verifying this is beyond the scope of this paper.
A more detailed discussion of Eq. \ref{eq:theorycurrent} and its derivation is given in Ref. \cite{singh2020josephson}.

\begin{thebibliography}{10}
\bibitem{idealjunction}
We note that achieving a sinusoidal current phase relation requires a barrier
  deep in the tunnelling regime which is fulfilled for our barrier depth of
  $V_0/\mu \approx 1.5$.
\end{thebibliography}
\end{document}